\documentclass[10pt, final, conference, letterpaper]{IEEEtran}
\usepackage[left=.6in,right=.6in,top=.719in,bottom=.99in]{geometry}

\usepackage{amsmath,amssymb,dsfont,stfloats,color,url,bbm}
\usepackage[pdftex]{graphicx}
\usepackage[caption=false,font=footnotesize]{subfig}
\usepackage{mathabx}
\usepackage{multirow}
\usepackage{tabu}

\DeclareGraphicsExtensions{.eps,.pdf,.png,.jpg,.gif,.jpeg,.pstex}

\usepackage[noadjust]{cite}
\usepackage{url}

\newfont{\bbb}{msbm10 scaled 700}

\newfont{\bb}{msbm10 scaled 1100}
\newcommand{\CC}{\mbox{\bb C}}

\newcommand{\EE}{\mbox{\bb E}}

\newcommand{\HH}{\mbox{\bb H}}

\newcommand{\yy}{\mathbbm{y}}

\newcommand{\zz}{\mathbbm{z}}
\newcommand{\sss}{\mathbbm{s}}

\newcommand{\hh}{\mathbbm{h}}

\newcommand{\vvv}{\mathbbm{v}}

\newcommand{\hv}{{\bf h}}

\newcommand{\rv}{{\bf r}}

\newcommand{\wv}{{\bf w}}
\newcommand{\vv}{{\bf v}}

\newcommand{\yv}{{\bf y}}
\newcommand{\zv}{{\bf z}}
\newcommand{\zerov}{{\bf 0}}

\newcommand{\Fm}{{\bf F}}

\newcommand{\Id}{{\bf I}}

\newcommand{\Ym}{{\bf Y}}
\newcommand{\Zm}{{\bf Z}}

\newcommand{\Cc}{{\cal C}}

\newcommand{\Ec}{{\cal E}}

\newcommand{\Gc}{{\cal G}}

\newcommand{\Nc}{{\cal N}}

\newcommand{\Sc}{{\cal S}}

\newcommand{\Uc}{{\cal U}}

\newcommand{\nuv}{\hbox{\boldmath$\nu$}}

\newcommand{\phiv}{\hbox{\boldmath$\phi$}}

\newcommand{\Sigmam}{\hbox{\boldmath$\Sigma$}}

\newcommand{\eqdef}{\stackrel{\Delta}{=}}

\newcommand{\herm}{{\sf H}}

\newcommand{\SINR}{{\sf SINR}}
\newcommand{\SNR}{{\sf SNR}}

\newcommand{\taudmrs}{\tau_p}

\usepackage{stmaryrd} 

\begin{document}

\setlength{\abovedisplayskip}{1pt}
\setlength{\belowdisplayskip}{1pt}
\setlength{\abovedisplayshortskip}{1pt}
\setlength{\belowdisplayshortskip}{1pt}

\title{Optimal User Load and Energy Efficiency in User-Centric Cell-Free Wireless Networks}

\author{\IEEEauthorblockN{Fabian G\"ottsch\IEEEauthorrefmark{1},
		Noboru Osawa\IEEEauthorrefmark{2}, Takeo Ohseki\IEEEauthorrefmark{2}, Kosuke Yamazaki\IEEEauthorrefmark{2}, Giuseppe Caire\IEEEauthorrefmark{1}}
	\IEEEauthorblockA{\IEEEauthorrefmark{1}Technical University of Berlin, Germany\\
		\IEEEauthorrefmark{2}KDDI Research Inc., Japan\\
		Emails: \{fabian.goettsch, caire\}@tu-berlin.de, \{nb-oosawa, ohseki, ko-yamazaki\}@kddi-research.jp}}

\maketitle


\begin{abstract}
Cell-free massive MIMO is a variant of multiuser MIMO and massive MIMO, in which the total number of antennas $LM$ is distributed among the $L$ remote radio units (RUs) in the system, enabling macrodiversity and joint processing. 
Due to pilot contamination and system scalability, each RU can only serve a limited number of users. Obtaining the optimal number of users simultaneously served on one resource block (RB) by the $L$ RUs regarding the sum spectral efficiency (SE) is not a simple challenge though, as many of the system parameters are intertwined. For example, the dimension $\tau_p$ of orthogonal Demodulation Reference Signal (DMRS) pilots limits the number of users that an RU can serve. Thus, depending on $\tau_p$, the optimal user load yielding the maximum sum SE will vary. 
Another key parameter is the users' uplink transmit power $P^{\rm ue}_{\rm tx}$, where a trade-off between users in outage, interference and energy inefficiency exists.
We study the effect of multiple parameters in cell-free massive MIMO on the sum SE and user outage, as well as the performance of different levels of RU antenna distribution.
We provide extensive numerical investigations to illuminate the behavior of the system SE with respect to the various parameters, including the effect of the system load, i.e., the number of active users to be served on any RB.
The results show that in general a system with many RUs and few RU antennas yields the largest sum SE, where the benefits of distributed antennas reduce in very dense networks.
\end{abstract}

\section{Introduction} 
Multiuser MIMO and massive MIMO have been key transformative technologies under research in the area of mobile communication networks throughout the last years, both the theoretical foundations and the practical system design \cite{caire2003achievable, 3gpp38211, marzetta2010noncooperative}. Massive MIMO is based on the fact that $M$ remote radio units (RUs) can be trained by $K$ user equipments (UEs) transmitting an uplink (UL) pilot signal of dimension $\taudmrs\geq K$. However, if $K>\taudmrs$ are served simultaneously on the same channel coherence block, pilot contamination occurs and creates a coherent combined term in the inter-cell interference that does not vanish as $M \rightarrow \infty$ \cite{marzetta2010noncooperative}. A promising approach to deal with pilot contamination is cell-free massive MIMO, where multiple RUs cooperate to jointly improve the communication quality of each user. 
In \cite{goettsch2021impact, kddi_uldl_precoding}, we have shown that pilot contamination can be reduced significantly with subspace projection channel estimation, yielding a system performance close to the case of ideal partial channel state information, where each remote radio unit (RU) has perfect knowledge of the channel vectors of its associated UEs. In these works, we assumed a system with $K=100$ UEs and different antenna distributions, i.e., different numbers of RUs $L$ and RU antennas $M$, where $LM \gg K$.
Eliminating pilot contamination however becomes more and more difficult with a growing number of users, as pilots are reused by more users leading to more severe pilot contamination, which in turn limit the system performance. 

While we aimed to maximize the sum spectral efficiency (SE) using different combining and precoding schemes in \cite{goettsch2021impact, kddi_uldl_precoding}, another direction of research focuses on the energy efficiency (EE), i.e., the sum data rate of all UEs divided by the total power consumption of the network at the UEs, RUs, and centralized processing units \cite{8097026}. 
The downlink (DL) EE is optimized in \cite{8097026}, where the optimization variables are the DL transmit (Tx) powers at the RUs for every UE. 
An extensive study of the DL EE in terms of various system parameters such as the RU density, the pilot reuse factor and the number of RU antennas and users, respectively, is provided in \cite{9353695}.
The UL EE is studied in \cite{8781848} with local maximum ratio combining (MRC) at the RUs, allowing the computation of closed-form expressions but yielding a degraded system performance compared to linear Minimum Mean-Square Error (LMMSE) combining.

In this paper, we investigate the UL of scalable user-centric cell-free wireless systems using local LMMSE combining in terms of UE density and UE transmit power for different levels of antenna distribution. Different to the previously mentioned works, we do not directly optimize the EE, but we investigate the sum SE for different UL transmit powers $P^{\rm ue}$, which is the same for all UEs. 
As the number of UEs in a wireless system does generally not change much in short time periods, the results give an idea of the UL transmit power that maximizes the sum SE without wasting energy and employing an algorithm for EE optimization. 
Another contribution of this paper is the investigation of the optimal user load and the achieved sum SE for different system configurations, where no UE is allowed to be in outage. We fix the number $LM$ of total RU antennas in the system and study the number of UEs $K$ and the pilot dimension $\taudmrs$ that yield the largest sum SE for different realizations of $L$ and $M$.
Furthermore, keeping $L$ and $K$ fixed, we consider different levels of RU and UE density, i.e., we vary the size of the network area. 

The simulation results show that optimizing the system performance of cell-free wireless networks is not straightforward, since many system parameters are interdependent in the sense that changing one parameter will lead to a change of another variable for the maximum system performance. Interesting insights for the design of cell-free wireless networks are obtained for various system aspects such as the RU and UE density, the pilot dimension, and the UE transmit power.


\section{System model}
We consider a cell-free wireless network in TDD operation with $L$ RUs, each with $M$ antennas, and $K$ single-antenna UEs. The set $\Cc_k \subseteq [L] = \{1,2,\dots, L\}$ denotes the cluster of RUs serving UE $k$ and the set $\Uc_\ell \subseteq [K]$ the cluster of UEs connected to RU $\ell$, where the cluster formation and DMRS pilot assignment for channel estimation follow the schemes in \cite{goettsch2021impact}. The RU-UE associations are described by a bipartite graph $\Gc$ whose vertices are the RUs and UEs, respectively. The set of edges accounting for associated RU-UE pairs is denoted by $\Ec$, i.e., $\Gc = \Gc([L], [K], \Ec)$. We consider OFDM modulation and channels following the standard block-fading model \cite{marzetta2010noncooperative,9336188,9064545}, such that they are random but constant over coherence blocks of $T$ signal dimensions in the time-frequency domain. 

We denote by $\HH \in \CC^{LM \times K}$ the channel matrix between all $LM$ RU antennas and all $K$ UE antennas on a given RB, where the $M \times 1$ block $\hv_{\ell,k}$ in the position corresponding to RU $\ell$ and UE $k$ denotes the channel between this RU-UE pair. We consider the {\em ideal partial CSI} regime, where each RU has perfect channel knowledge for its associated UEs. The known channel matrix of a cluster $\Cc_k$ is denoted by $\HH(\Cc_k) \in \CC^{LM \times K}$,
whose $M \times 1$ blocks of RU-UE pairs $(\ell,j) \in \Ec$ with $\ell \in \Cc_k$ are equal to $\hv_{\ell,j}$, and equal to $\zerov$ otherwise. 
The indiviudal channels between RUs and UEs follow the simplified single ring local scattering model \cite{adhikary2013joint}, and $\Fm$ denotes the $M \times M$ unitary DFT matrix with $(m,n)$-elements
$\Fm_{m,n} = \frac{e^{-j\frac{2\pi}{M} mn}}{\sqrt{M}}$ for  $m, n  = 0,1,\ldots, M-1$.
Then, the channel between RU $\ell$ and UE $k$ is
\begin{equation} 
	\hv_{\ell,k} = \sqrt{\frac{\beta_{\ell,k} M}{|\Sc_{\ell,k}|}}  \Fm_{\ell,k} \nuv_{\ell, k},
\end{equation}
where $\Sc_{\ell,k} \subseteq \{0,\ldots, M-1\}$, $\nuv_{\ell,k}$ and $\beta_{\ell,k}$ are the angular support set according to \cite{adhikary2013joint}, an $|\Sc_{\ell,k}| \times 1$ i.i.d. Gaussian vector with components 
$\sim \Cc\Nc(0,1)$, and the LSFC including
pathloss, blocking effects and shadowing, respectively. 
Using a Matlab-like notation, $\Fm_{\ell,k} \eqdef \Fm(: , \Sc_{\ell,k})$ denotes the tall unitary matrix obtained by selecting the columns of $\Fm$ corresponding to the index set $\Sc_{\ell,k}$.

\subsection{Uplink data transmission} 
Let all UEs transmit with the same average energy per symbol $P^{\rm ue}$ (usually referred to as “power”), and we define the system parameter 
\begin{equation}
	\SNR \eqdef P^{\rm ue}/N_0, \label{eq_snr}
\end{equation}
where $N_0$ denotes the complex baseband noise power spectral density.
We let $\sss^{\rm ul} \in \CC^{K \times 1}$  denote the vector of modulation symbols transmitted collectively by the $K$ users over the UL at a given symbol time (channel use).
The symbols $s_k^{\rm ul}$ ($k$-th components of $\sss^{\rm ul}$ are i.i.d. with mean zero and unit variance. 
The observation (samples) at the $LM$ RUs antennas for a single UL channel use is given by 
\begin{equation} 
	\yy^{\rm ul} = \sqrt{\SNR} \; \HH \sss^{\rm ul}   + \zz^{\rm ul}, \label{ULchannel}
\end{equation}
where $\zz^{\rm ul}$ has i.i.d. components $\sim \Cc\Nc(0,1)$. 
The goal of the cluster $\Cc_k$ processor  is to produce an estimate of 
$s^{\rm ul}_k$ which is treated by the channel decoder for user $k$ as the soft-output of a virtual single-user channel.  
As in most works on massive/cell-free MIMO, we restrict the cluster processing to be linear and 
define the receiver unit-norm vector $\vvv_k \in \CC^{LM \times 1}$ formed by blocks
$\vv_{\ell,k}  \in \CC^{M \times 1}, \ \ell \in [L]$, such that $\vv_{\ell,k} = \zerov$ (the identically zero vector) if $(\ell,k) \notin \Ec$. 
The non-zero blocks $\vv_{\ell,k} :  \ell \in \Cc_k$ are defined according to a local LMMSE principle (to be detailed later).

The corresponding scalar combined observation for symbol $s^{\rm ul}_k$ is given by 
\begin{eqnarray}
	r^{\rm ul}_k  & = & \vvv_k^\herm \yy^{\rm ul} \nonumber \\
	& = & \sqrt{\SNR} \vvv_k^\herm \hh_k s^{\rm ul}_k   + \sqrt{\SNR} \vvv_k^\herm \HH_k \sss_k^{\rm ul}  + \vvv_k^\herm \zz^{\rm ul} ,  \label{received-UL}
\end{eqnarray}
where we let $\hh_k$ denote the $k$-th column of $\HH$, where $\HH_k$ is the matrix $\HH$ after elimination of the $k$-th column, 
and $\sss_k^{\rm ul} $ is the vector $\sss^{\rm ul}$ after elimination of the $k$-th element. 

The resulting UL {\em instantaneous} SINR is given by 
\begin{eqnarray} 
	\SINR^{\rm ul}_k  & = & \frac{  |\vvv_k^\herm \hh_k|^2 }{ \SNR^{-1}  + \sum_{j \neq k} |\vvv_k^\herm \hh_j |^2 }.  \label{UL-SINR-unitnorm}
\end{eqnarray}
As a performance measure, in this paper we consider the so-called {\em optimistic ergodic rate}, given by 
\begin{eqnarray}
	R^{\rm ul}_k = \EE [ \log (1 + \SINR^{\rm ul}_k) ], \label{ergodic_rate_ul}
\end{eqnarray}
where the expectation is with respect to the small scale fading,
while conditioning on the long-term variables (LSFCs, placement of UEs and RUs on the plane, and cluster formation).

\section{Uplink local LMMSE detection}
Each RU $\ell$ makes use of locally computed receiving vectors $\vv_{\ell,k}$ for 
its users $k \in \Uc_\ell$. 
Letting 
\begin{eqnarray}
	\yv_\ell^{\rm ul} & =  & \sqrt{\SNR} \sum_{j=1}^K \hv_{\ell,j} s_j^{\rm ul}   + \zv_\ell^{\rm ul},  
	\label{receivedsignal-ell}
\end{eqnarray}
the $M \times 1$ block of $\yy^{\rm ul}$ corresponding to RU $\ell$,  RU $\ell$ computes the local detector
$r^{\rm ul}_{\ell,k} = \vv_{\ell,k}^\herm \yv_\ell^{\rm ul}$ for each $k \in \Uc_\ell$ and sends
the symbols $\{r^{\rm ul}_{\ell,k} : k \in \Uc_\ell\}$ to the decentralized units (DUs) implementing cluster-based combining. Then, the processor for cluster $\Cc_k$ (implemented at some DU)
computes the cluster-level combined symbol 
\begin{equation} 
	r^{\rm ul}_k = \sum_{\ell \in \Cc_k} w^*_{\ell,k} r^{\rm ul}_{\ell,k} = \wv^\herm_k \rv^{\rm ul}_k,   \label{combining_perfect_csi}
\end{equation}
where we define $\rv^{\rm ul}_k = \{ r^{\rm ul}_{\ell,k} : \ell \in \Cc_k\}$ and where $\wv_k$ is a vector of combining coefficients. 
We consider the case where the $\vv_{\ell,k}$'s are the LMMSE receiver vectors given the partial local CSI at RU $\ell$, while treating the out of cluster interference component (whose CSI is unknown to RU $\ell$) as an increase of the variance of the white Gaussian noise term, 
given by (see details in  \cite{goettsch2021impact}) 
\begin{equation} 
	\sigma^2_\ell =  1 + \SNR \sum_{j \neq \Uc_\ell}  \beta_{\ell,j}. \label{sigmaell} 
\end{equation}
Under this assumption, the LMMSE receiving vector for user $k$ at RU $\ell$ is given by 
\begin{equation} 
	\vv_{\ell,k} = \left ( \sigma_\ell^2 \Id + \SNR \sum_{j \in \Uc_\ell} \hv_{\ell,j} \hv_{\ell,j}^\herm \right )^{-1} \hv_{\ell,k}.  \label{eq:lmmse}
\end{equation}
The vector of combining coefficients $\wv_k$ is computed to maximize a {\em nominal SINR} expression, i.e., the SINR under the assumption that the out-of-cluster interference is replaced by a corresponding white noise term with the same per-component variance \cite{goettsch2021impact}.

\section{UL channel estimation}
In practice, ideal partial CSI is not available and the channels $\{\hv_{\ell,k} : (\ell,k) \in \Ec\}$ must be estimated from UL pilots. 
We assume that $\taudmrs$ signal dimensions per slot are dedicated to UL pilots and define a codebook of $\taudmrs$ orthogonal pilot 
sequences.  The pilot field received at RU $\ell$ is given by the $M \times \taudmrs$ matrix 
\begin{equation} 
	\Ym_\ell^{\rm pilot} = \sum_{i=1}^K \hv_{\ell,i} \left(\phiv_{t_i}\right)^\herm + \Zm_\ell^{\rm pilot} \label{Y_pilot}
\end{equation}
where $\Zm_\ell^{\rm DMRS}$ is AWGN with elements i.i.d. $\sim \Cc\Nc(0, 1)$, and $\phiv_{t_i}$ denotes the pilot  vector of dimension $\taudmrs$ used by UE $i$ at the current slot, normalized such that
$\| \phiv_{t_i} \|^2 = \taudmrs\SNR$ for all $t_i \in [\taudmrs]$. The standard {\em Least-Squares} channel estimation used in overly many papers on massive MIMO
consists of ``pilot matching'', i.e., RU $\ell$ produces the channel estimates \begin{gather} 
	\widehat{\hv}^{\rm pm}_{\ell,k} = \frac{1}{\taudmrs \SNR} \Ym^{\rm pilot}_\ell \phiv_{t_k}  
	= \hv_{\ell,k}  + \sum_{i : t_i = t_k} \hv_{\ell,i}  + \widetilde{\zv}_{t_k,\ell}   \label{chest}
\end{gather} 
by right-multiplication of the pilot field by $\phiv_{t_k}$  for all $k \in \Uc_\ell$,  where $\widetilde{\zv}_{t_k,\ell}$ is $M \times 1$ Gaussian i.i.d. with components $\Cc\Nc(0, \frac{1}{\taudmrs\SNR})$ and where  $\sum_{i : t_i = t_k} \hv_{\ell,i}$ is the pilot contamination term, i.e., the contribution of the channels from users $i \neq \Uc_\ell$ using the same pilot $t_k$ (co-pilot users).  Assuming that the subspace information $\Fm_{\ell,k}$ of all $k \in \Uc_\ell$ is known, we consider the ``subspace projection'' (SP) 
pilot decontamination scheme given by the orthogonal projection of $\widehat{\hv}^{\rm pm}_{\ell,k}$ onto the subspace spanned by the columns of 
$\Fm_{\ell,k}$, i.e., 
\begin{align}
	\widehat{\hv}^{\rm sp}_{\ell,k} & = \Fm_{\ell,k}\Fm_{\ell,k}^\herm \widehat{\hv}^{\rm pm}_{\ell,k} \nonumber \\
	& = \hv_{\ell,k}  + \Fm_{\ell,k}\Fm_{\ell,k}^\herm \left ( \sum_{i : t_i = t_k} \hv_{\ell,i} \right ) + \Fm_{\ell,k} \Fm_{\ell,k}^\herm \widetilde{\zv}_{t_k,\ell}.   \label{chest1}
\end{align}
Writing explicitly the pilot contamination term after the subspace projection, it is immediate to show that its covariance matrix is given by 
\begin{equation}
	\Sigmam_{\ell,k}^{\rm co}  = \sum_{i : t_i = t_k} \frac{\beta_{\ell,i} M}{|\Sc_{\ell,i}|}  \Fm_{\ell,k} \Fm_{\ell,k}^\herm \Fm_{\ell,i} \Fm^\herm_{\ell,i} \Fm_{\ell,k} \Fm^\herm_{\ell,k} .
\end{equation}
When $\Fm_{\ell,k}$ and $\Fm_{\ell,i}$ are nearly mutually orthogonal, i.e., $\Fm_{\ell,k}^\herm \Fm_{\ell.i} \approx \zerov$,
the subspace projection is able to significantly reduce the pilot contamination effect. 

\section{Simulations} \label{sec:ul_perfect_csi}
In our system simulations, we assume non-ideal CSI by replacing the ideal partial CSI $\{ \hv_{\ell,k} : (\ell,k) \in \Ec \}$ with the estimated partial CSI $\{ \widehat{\hv}^{\rm sp}_{\ell,k} : (\ell,k) \in \Ec \}$ in (\ref{eq:lmmse}). We consider a square coverage area with a torus topology to avoid boundary effects. 
The LSFCs are given  according to the 3GPP urban microcell pathloss model from \cite{3gpp38901}. 
We consider RBs of dimension $T = 200$ symbols. 
The UL spectral efficiency (SE) for UE $k$ is given by  
\begin{equation}
	{\rm SE}^{\rm ul}_k =  (1 - \taudmrs/T) R_k^{\rm ul}.
\end{equation}
The angular support $\Sc_{\ell,k}$ contains the DFT quantized angles (multiples of $2\pi/M$) falling inside an interval of length $\Delta$ placed symmetrically around the direction joining UE $k$ and RU $\ell$. We use $\Delta = \pi/8$ and the maximum cluster size $Q=10$ (RUs serving one UE) in the simulations. The SNR threshold $\eta=1$ makes sure that an RU-UE association can only be established, when $\beta_{\ell,k} \geq \frac{\eta}{M \SNR } $. We study several systems configurations with different levels of antenna distribution, where the total number of RU antennas is $LM=640$ for all configurations.

We consider a bandwidth of $W = 10\text{ MHz}$ and noise with power spectral density of $N_0 = -174 \text{ dBm/Hz}$. The UL energy per symbol $P^{\rm ue}$ in (\ref{eq_snr}) is one of the parameters that we will study, since the system performance does not necessarily improve with larger $P^{\rm ue}$ due to interference. 
The actual (physical) UE transmit power
is obtained as $P_{\rm tx}^{\rm ue} = P^{\rm ue}W$ and it is expressed in dBm.  For each set of parameters, we generated 50 independent layouts, and for each layout we computed the expectation in (\ref{ergodic_rate_ul}) by Monte Carlo averaging with respect to the channel vectors.

\subsection{Comparison of RU antenna distribution levels} \label{sec:extensive_simulations}
\begin{table}[htbp]
	\caption{Studied parameters and their respective values.}
	 \vspace{-.4cm}
	\begin{center}
		\begin{tabular}{ | c | c |}
			\hline
			{ Parameter } & { Values} \\ \hline
			$L$ & \{10,20,40,80,160\} \\ \hline
			$K$ & \{100, 300, 500, 750, 1000, 1250, 1500\} \\ \hline
			$\tau_p$ & \{10,20,30,40,50,60,80,100,125,150\} \\ \hline
			$P_{\rm tx}^{\rm ue}$ & \{-10, 0, 10, 20\} dBm \\ \hline
		\end{tabular}
		\label{table_params}
	\end{center}
\end{table}
\vspace{-.6cm}
\begin{table}[htbp]
	\caption{Studied parameters and the values that achieve the largest sum SE in each setup, where $P_{\rm tx}^{\rm ue}$ and $L$ are fixed.}
	\vspace{-.4cm}
	\begin{center}
		\begin{tabu}{ | c | c | c | c | c |}
			\hline
			{ $P_{\rm tx}^{\rm ue}$ } & {$L$}  & {$K$}  & {$\tau_p$} & Sum SE [bit/s/Hz] \\ 
			[-0pt] \tabucline[1.5pt]{1-5}
			\multirow{5}{*}{$-10$ dBm} & 10 & 1250 & 60 & 895.0 \\  
			\cline{2-5} & 20 & 1500 & 50 & 1067.2 \\ 
			\cline{2-5} & 40 & 1500 & 40 & 1133.6 \\
			\cline{2-5} & 80 & 1250 & 20 & 1144.1 \\  
			\cline{2-5} & 160 & 1250 & 20 & 1107.3 \\  
			[-1.5pt] \tabucline[1.5pt]{1-5}
			\multirow{5}{*}{$0$ dBm} & 10 & 500 & 50 & 810.2 \\  
			\cline{2-5} & 20 & 750 & 50 & 1029.1 \\ 
			\cline{2-5} & 40 & 1000 & 40 & 1141.1 \\ 
			\cline{2-5} & 80 & 750 & 20 & 1174.1 \\  
			\cline{2-5} & 160 & 750 & 20 & 1142.6 \\  
			[-1.5pt] \tabucline[1.5pt]{1-5}
			\multirow{5}{*}{$10$ dBm} & 10 & 500 & 60 & 795.1 \\  
			\cline{2-5} & 20 & 750 & 50 & 1038.8 \\ 
			\cline{2-5} & 40 & 750 & 30 & 1145.7 \\ 
			\cline{2-5} & 80 & 1000 & 20 & 1181.3 \\  
			\cline{2-5} & 160 & 750 & 20 & 1150.9 \\  
			[-1.5pt] \tabucline[1.5pt]{1-5}
			\multirow{5}{*}{$20$ dBm} & 10 & 300 & 50 & 819.4 \\  
			\cline{2-5} & 20 & 750 & 50 & 1030.3 \\ 
			\cline{2-5} & 40 & 750 & 40 & 1142.7 \\ 
			\cline{2-5} & 80 & 1000 & 20 & 1191.3 \\  
			\cline{2-5} & 160 & 750 & 20 & 1154.8 \\  \hline
		\end{tabu}
		\label{table_sumSE}
	\end{center}
\end{table}
We carry out extensive simulations with different values for $P_{\rm tx}^{\rm ue}$, $\tau_p$ and $K$ in systems with a total number of $LM=640$ RU antennas distributed on an area of $A = 2 \times 2$ $\text{km}^2$. The considered values for the parameters under investigation are summarized in Table \ref{table_params}. 
First, we want to get an overview which RU antenna configurations, i.e., the specific values of $L$ and $M$, lead to the best system performance with respect to the sum SE of all UEs. 
However, due to space limitations, we cannot show the achieved sum SE for every combination of the system parameters. 
Instead, for each considered value of $P_{\rm tx}^{\rm ue}$, we give in Table \ref{table_sumSE} the values of $K$ and $\tau_p$ that yield the maximum sum SE in each configuration of $L$ and $M$.
The more concentrated configurations achieve significantly smaller sum SEs for all transmit powers. This occurs since for smaller $L$, each RU needs to serve UEs in a larger coverage area on average. leading to smaller channel gains and decreased macrodiversity. 
Therefore, we will focus on the analysis of $L=\{40,80\}$ in the following. For $L=40$ and $L=80$, we consider $\tau_p=\{30,40\}$ and $\tau_p=\{20,30\}$, respectively, which are the parameters that yield the best system performance. A higher level of RU antenna distribution with $L=160$ did not yield any further improvement of the sum SE.

\begin{figure}[t!]
	\centerline{\includegraphics[width=.47\linewidth]{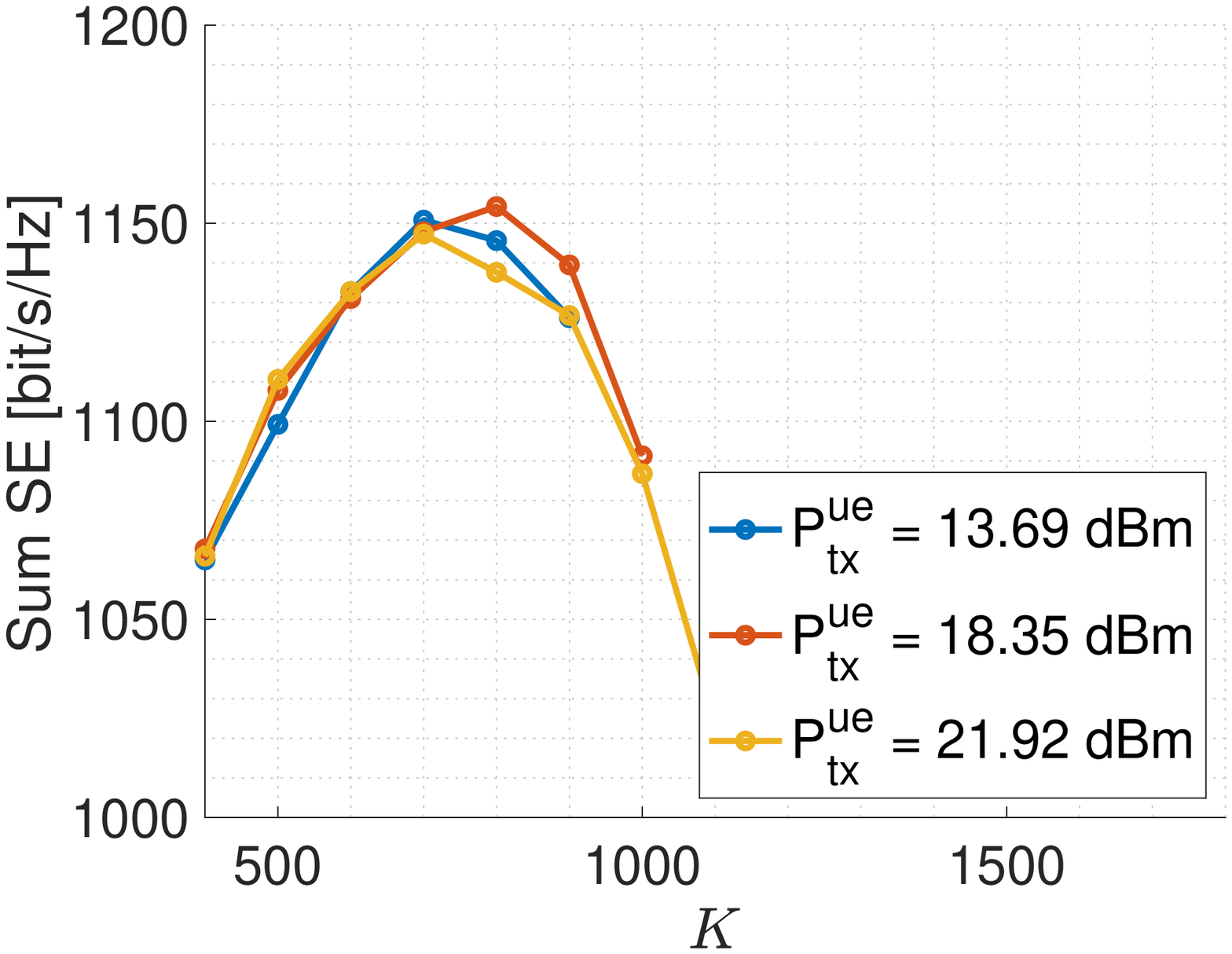} \hspace{.03\linewidth} \includegraphics[width=.47\linewidth]{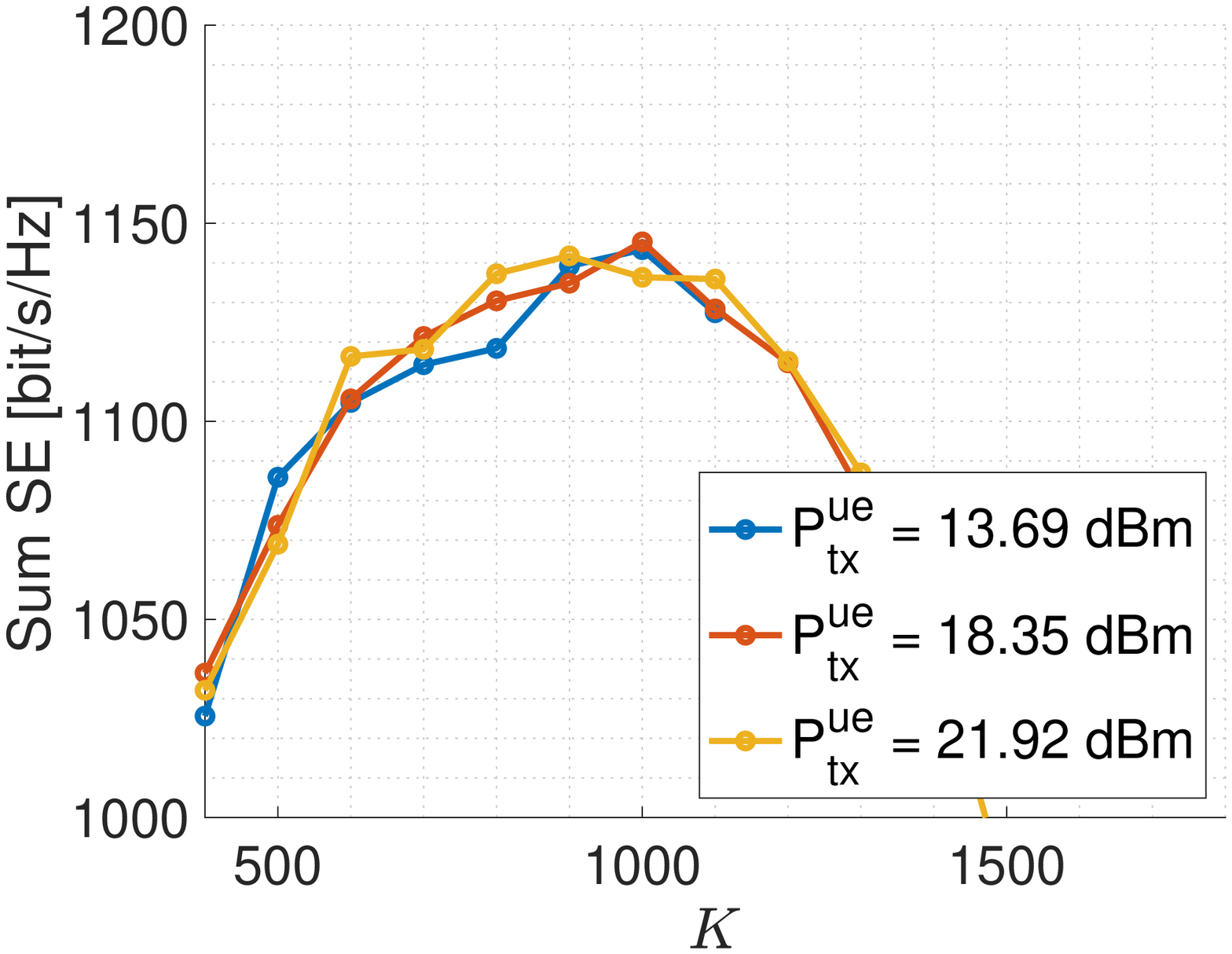}   }
	\vspace{-.3cm}
	\caption{Sum SE vs. $K$ for different $P_{\rm tx}^{\rm ue}$, where $L=40$. The left plot shows the results for $\tau_p = 30$, the right plot for $\tau_p = 40$.}
	\label{fig:sum_se_vs_K_L40}
\end{figure}

\begin{figure}[t!]
	\centerline{\includegraphics[width=.47\linewidth]{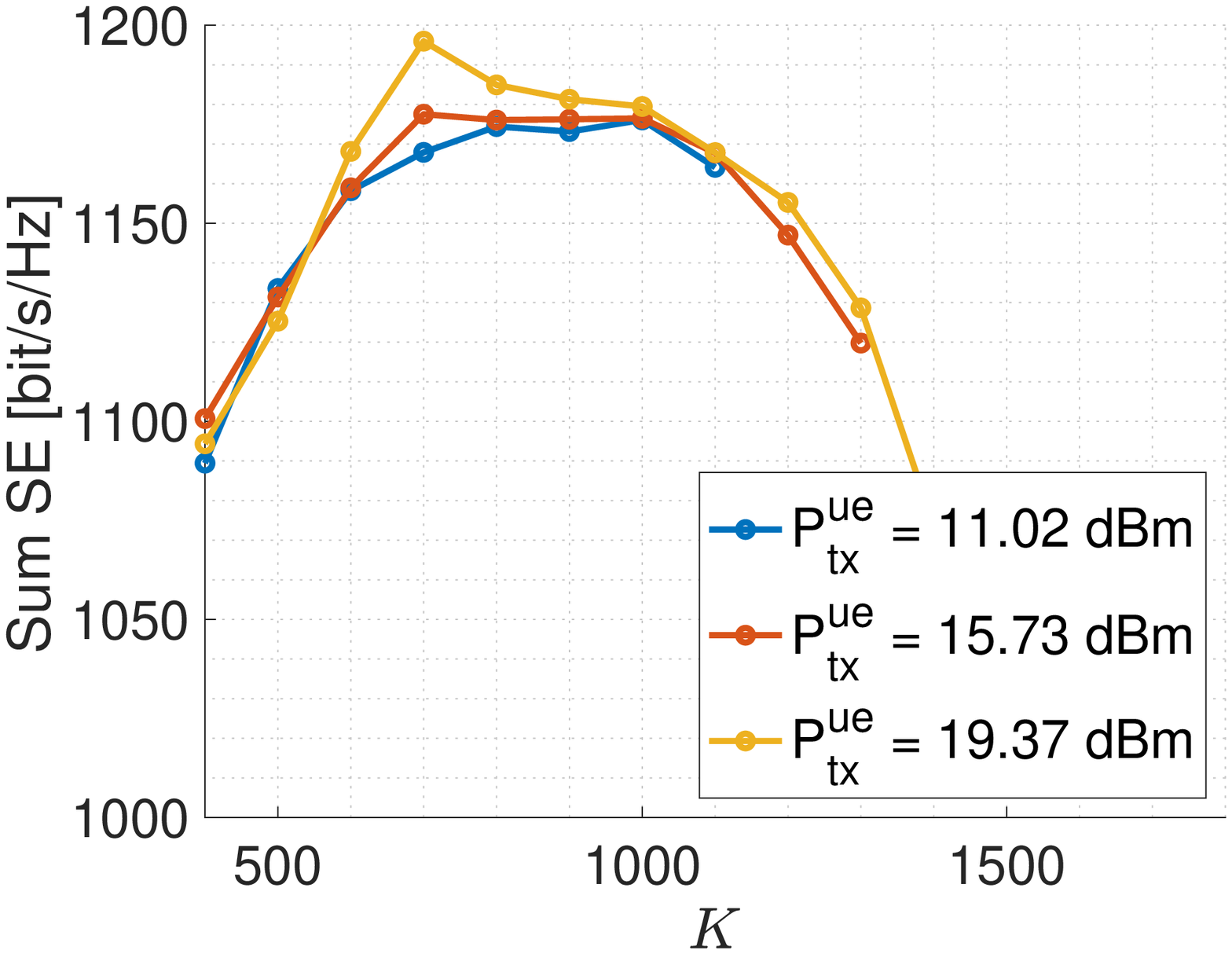} \hspace{.03\linewidth} \includegraphics[width=.47\linewidth]{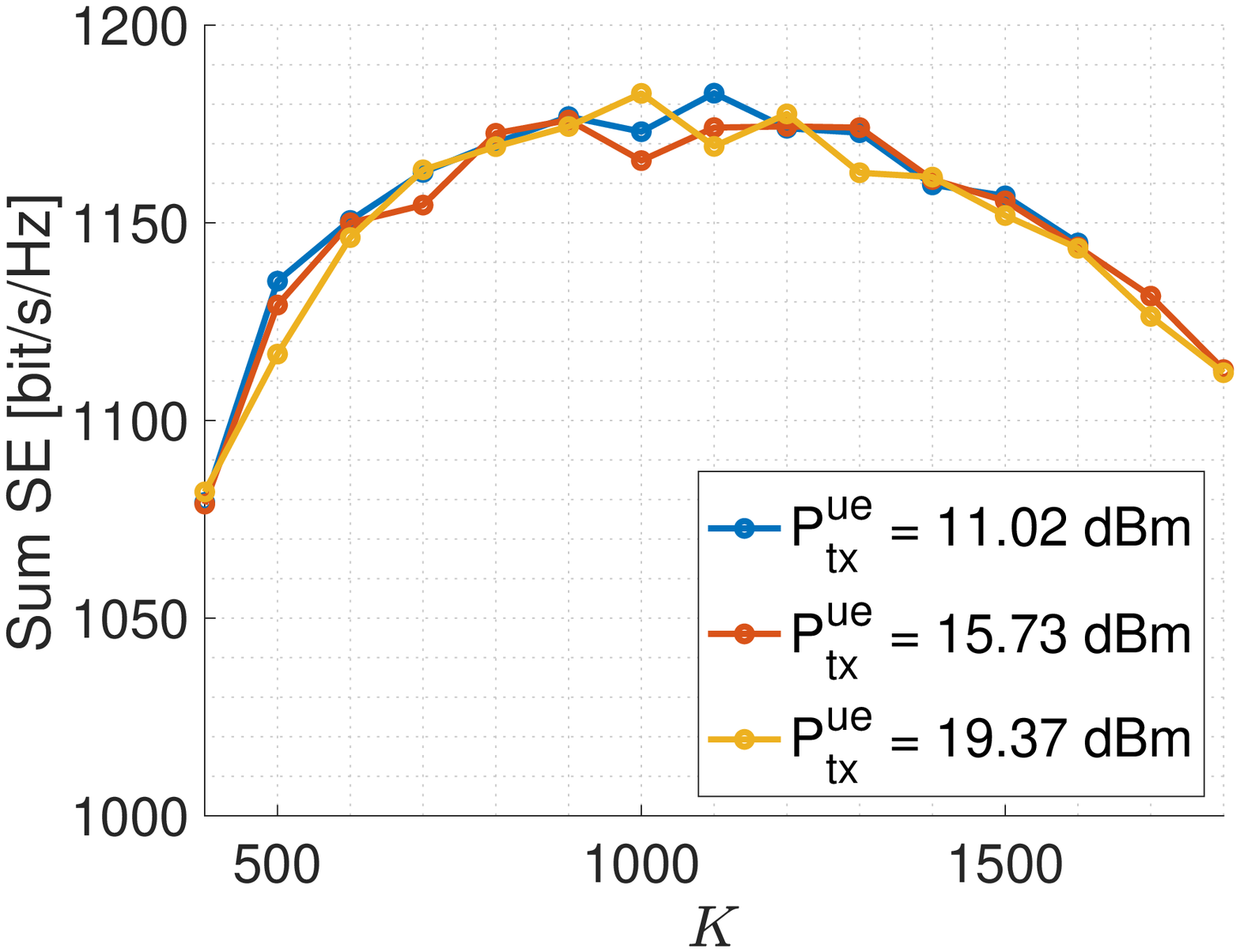}   }
	\vspace{-.3cm}
	\caption{Sum SE vs. $K$ for different $P_{\rm tx}^{\rm ue}$, where $L=80$. The left plot shows the results for $\tau_p = 20$, the right plot for $\tau_p = 30$.}
	\label{fig:sum_se_vs_K_L80}
\end{figure}

\subsection{Optimal user load and power efficiency}
In this section, we let the UL transmit power be dependent on the geometry and the number of RUs. The UL energy per symbol $P^{\rm ue}$ in (\ref{eq_snr}) is chosen such that $\bar{\beta} M \SNR = 1$ (i.e., 0 dB), when the expected pathloss $\bar{\beta}$ with respect to LOS and NLOS is calculated for distance 
$\rho d_L$, where $d_L = \sqrt{\frac{A}{\pi L}}$ is the radius of a disk of area equal to $A/L$. We let the UL Tx power of all UEs be dependent on the RU density and number of RU antennas to achieve a certain level of overlap of the RUs' coverage areas, such that each UE is likely to be associated to several RUs.
Notice that for all values of $L$, we observed a significant percentage of UEs in outage, i.e., UEs not served by any RU, for $P_{\rm tx}^{\rm ue} = \{-10, 0\}$ dBm. With $P_{\rm tx}^{\rm ue} = -10$ dBm and all considered $\tau_p$, more than 25\% of the UEs were in outage for $L=40$, and approximately 15\% for $L=80$. Since we are interested in the number of UEs simultaneously being served that maximizes the sum SE, we only consider system configurations in the following, where no UE is in outage. 

Figs. \ref{fig:sum_se_vs_K_L40} and \ref{fig:sum_se_vs_K_L80} show the sum SE with respect to $K$ for different values of $P_{\rm tx}^{\rm ue}$, where $\rho=\{3,4,5\}$. Note that only the data points of system configurations are plotted, for that no UE is in outage. We also simulated that the users transmit with $P_{\rm tx}^{\rm ue}$ using $\rho=\{1,2\}$, but in these cases the UE outage probability was non-zero.
When comparing $L=40$ and $L=80$, we observe that the configuration with $L=80$ outperforms $L=40$ with respect to the sum SE by a small margin in the order of tens of bit/s/Hz, due to the increased channel gains and macrodiversity as explained before.
As expected, more users can be supported for larger $\tau_p$, but not necessarily a larger sum SE is achieved. 
The number of UEs $K$ that yields the highest sum SE is bigger in case of $L=80$ RUs. We can explain this by the shorter distances between the RUs and UEs, and the number of potentially associated UEs to all RUs, i.e., $\taudmrs L$. The considered parameters, which optimize the sum SE as we have seen in section \ref{sec:extensive_simulations}, lead to $\taudmrs L= \{1600, 2400\}$ for $L=80$ and $\taudmrs = \{20,30\}$, respectively, while for $L=40$ the number of potentially associated UEs is $\taudmrs L = \{1200, 1600\}$.
The largest transmit power also does not lead to the largest sum SE in all cases, since it may cause a higher level of interference. When we consider that the difference of the achieved sum SE for the considered $P_{\rm tx}^{\rm ue}$ is relatively small (e.g., for $K$=1000, a sum SE difference of 20 bit/s/Hz would on average lead to a difference of 0.02 bit/s/Hz per UE), it may be more efficient to transmit with $11.02$ dBm than with $19.37$ dBm in the case $L=80$. 

\begin{figure}[t!]
	\centerline{\includegraphics[width=.49\linewidth]{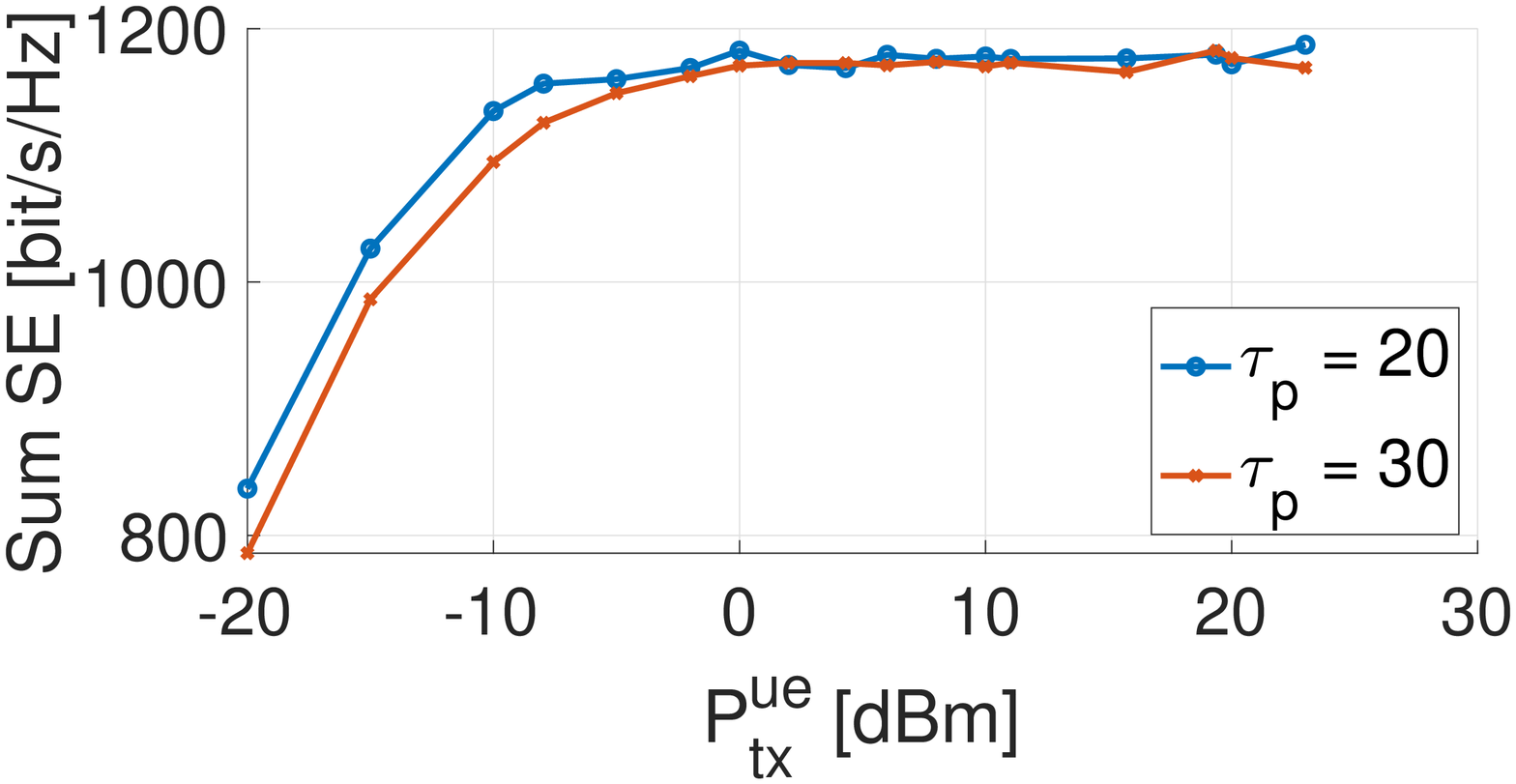} \includegraphics[width=.49\linewidth]{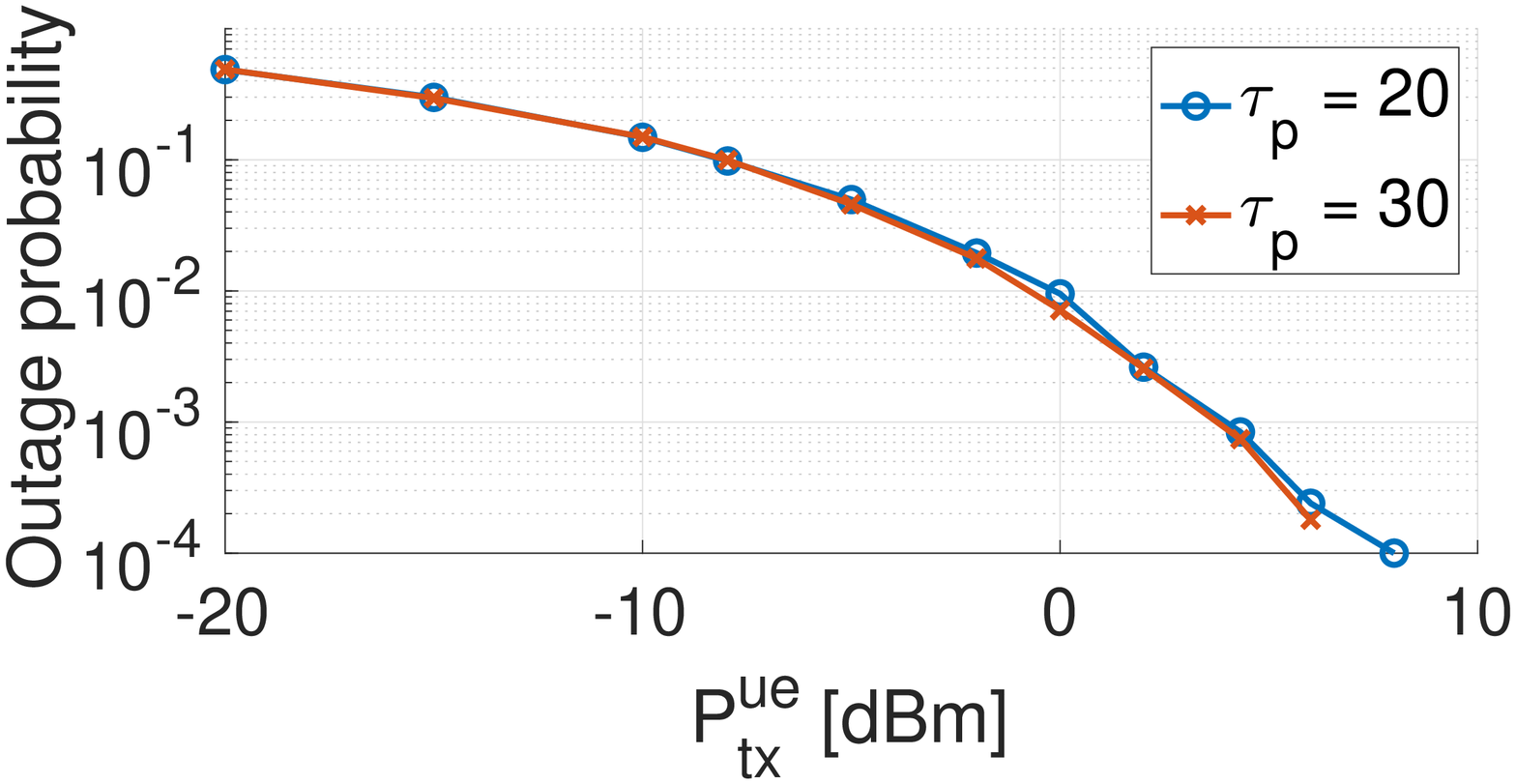}   }
	\vspace{-.3cm}
	\caption{System performance in terms of sum SE (left) and outage probability (right) for different $P_{\rm tx}^{\rm ue}$, where $L=80$ and $K=1000$.}
	\label{fig:sum_se_and_outage_vs_Ptx_L80}
\end{figure}

These results are confirmed by the simulation results in Fig. \ref{fig:sum_se_and_outage_vs_Ptx_L80}, where the sum SE and user outage probability are shown for different values of $P_{\rm tx}^{\rm ue}$ in a system with $L=80$, $K = 1000$ and $\tau_p = \{20, 30\}$. For $P_{\rm tx}^{\rm ue} < 10 {\rm dBm}$, the outage probability is non-zero, and grows for smaller transmit powers, where the difference between $\tau_p = 20$ and $\tau_p = 30$ is not significant. Only for $P_{\rm tx}^{\rm ue} = 8 {\rm dBm}$ and $\tau_p = 30$, the outage probability is zero, while it is non-zero for $\tau_p = 20$. The sum SE is basically unchanged for most values of $P_{\rm tx}^{\rm ue}>-10 \text{ dBm}$, with small variations. The fact that the sum SE does not vary much for $P_{\rm tx}^{\rm ue}>-10 \text{ dBm}$ can be explained as follows. Since all UEs transmit with low power and some UEs end up without connection to any RU, the signals of the connected UEs suffer from less interference. The RUs only connect to the UEs with the strongest channels, such that the sum SE is still large. For bigger $P_{\rm tx}^{\rm ue}$, the desired signal strength is larger, but also the interference, such that the sum SE saturates at some point, when $P_{\rm tx}^{\rm ue}$ is increased. For $P_{\rm tx}^{\rm ue} <-10 \text{ dBm}$ however, a smaller sum SE is achieved, since at some point many UEs are in outage and also the desired signal strength becomes too weak.

\subsection{Size of the network area and the pathloss effect}
\begin{figure}[t!]
	\centerline{\includegraphics[width=.49\linewidth]{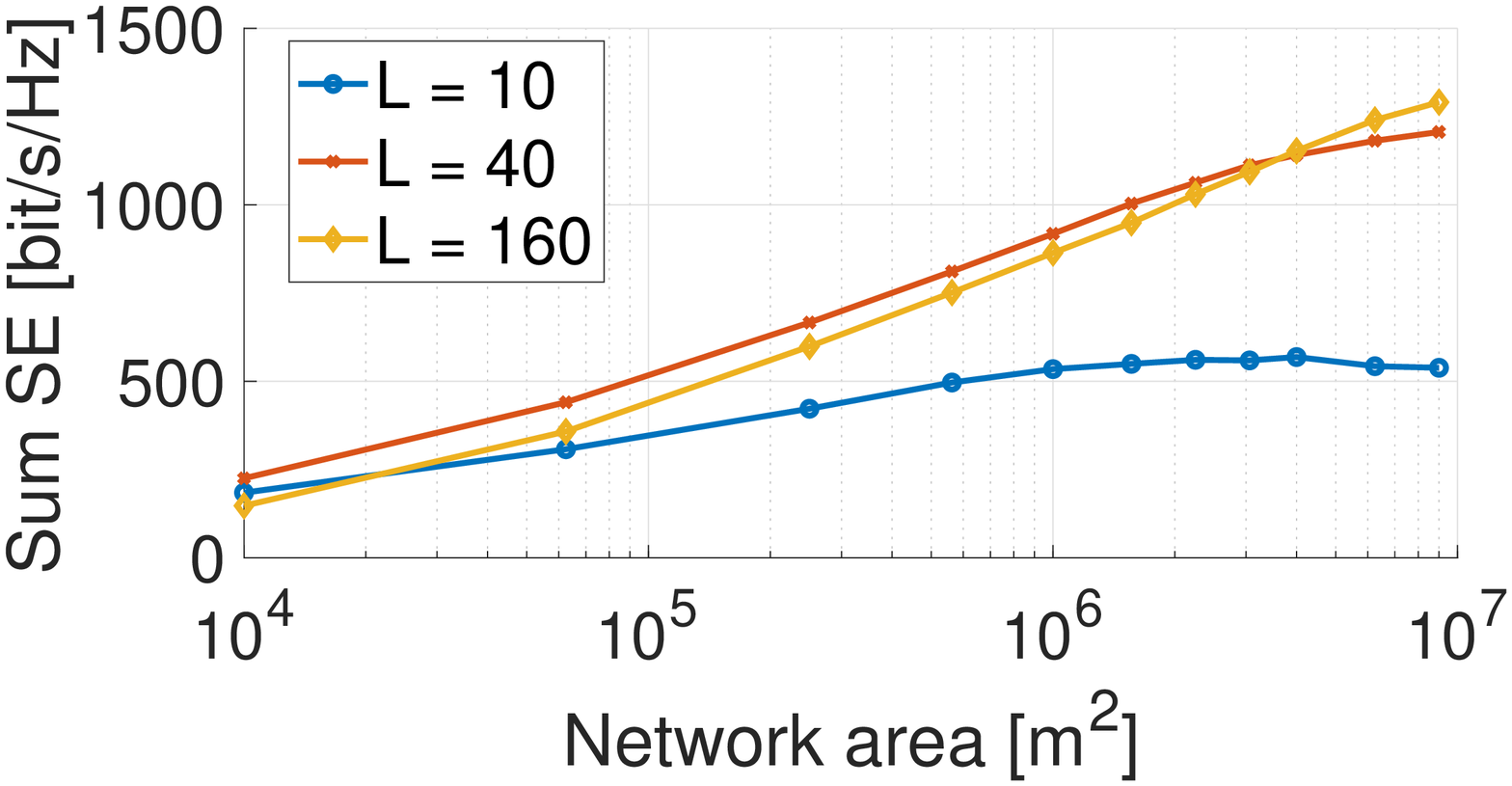}
	\includegraphics[width=.49\linewidth]{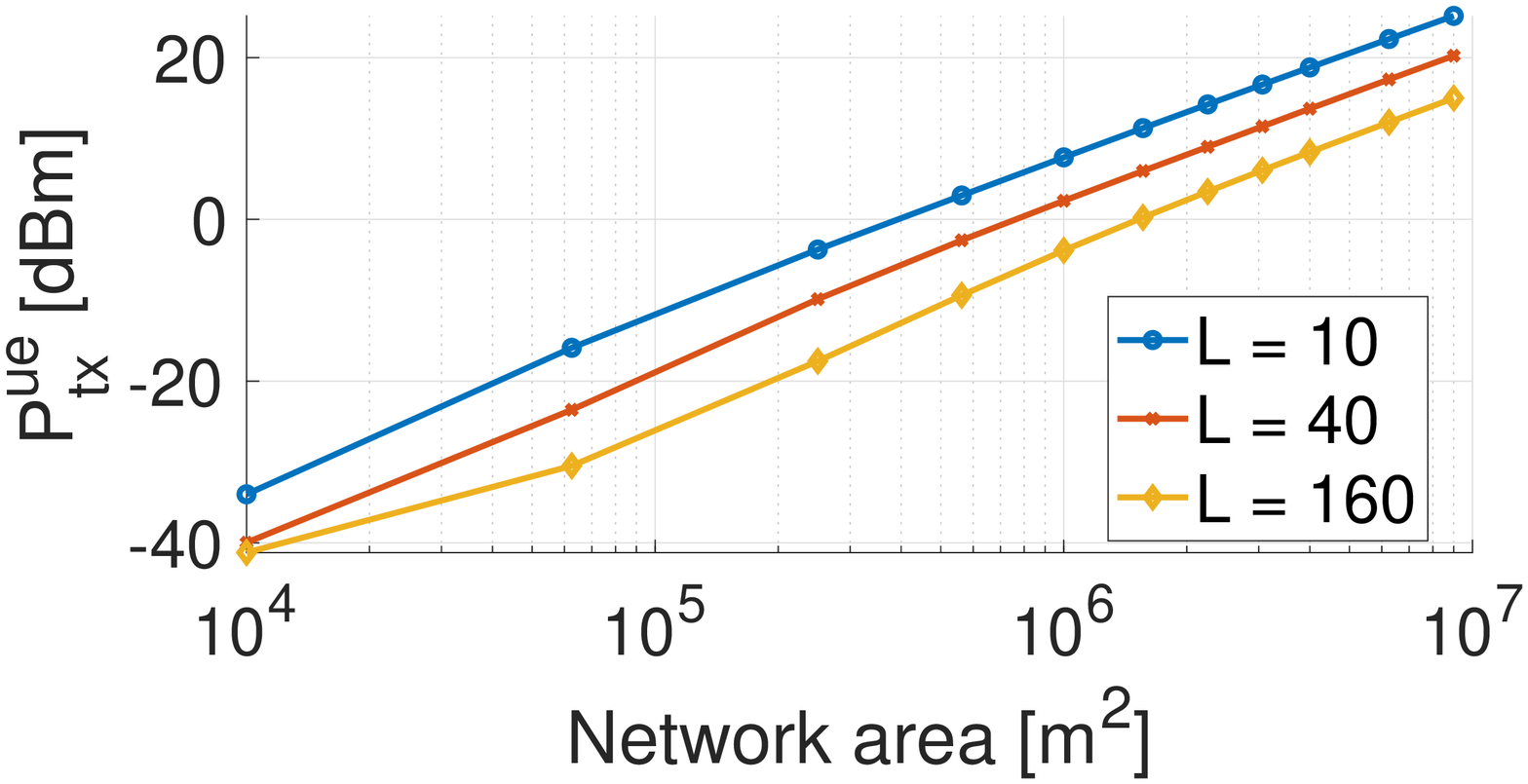}   
	}
	\vspace{-.3cm}
	\caption{Sum SE (left) and UL transmit power (right) vs. network area for $K = 1000$ and different $L$.}
	\label{fig:sum_se_vs_area}
\end{figure}
We conclude this section by considering squared areas with a different network area $A$ between $10^4$ and $10^7$ square meters. For a fair comparison, we let $K=1000$ UEs transmit according to the introduced geometry-based model with $\rho=3$ such that we have an overlap of the RU coverage areas, and consider different $L$ with $LM$ fixed. The parameter $\tau_p$ is optimized for each combination of $K$ and $L$ based on the extensive simulations discussed in section \ref{sec:extensive_simulations}.

Fig. \ref{fig:sum_se_vs_area} shows that a more concentrated antenna distribution such as $L=10$ can outperform a more distributed configuration like $L=160$ for small network areas $A$ (left plot). In general, the gap between the system performance of the configurations decreases for smaller networks. 
This can be explained by the fact that for a smaller network area the distances of RU-UE pairs do not differ a lot for the different choices of $L$. In a relatively large area, the distance between associated RUs and UEs may become very large for small $L$, while maintaining short distances (i.e., high channel gains) for large $L$. Then the benefits of increased channel gains and probability of LOS between RU-UE pairs in cell-free massive MIMO become obvious with a more distributed antenna configuration. However, this benefit becomes less significant if the network area decreases, and the array beamforming gain (increasing with $M$) at some point compensates for the larger distances between RU-UE pairs, when $L$ is small. For $L=10$, we can observe convergence of the sum SE, and conclude that a network area between $10^6$ and $10^7 \text{ m}^2$ corresponds to the optimal density of users and RUs for this configuration. The more distributed the antenna configuration, the larger network areas can be supported before the sum SE converges, thanks to increased macrodiversity. We expect a converging behavior of the sum SE also for $L=\{40,160\}$, when further increasing $A$.

The overall larger sum SE for geographically large networks is explained by the Tx power (see right plot of Fig. \ref{fig:sum_se_vs_area}) and the pathloss model.
The Tx power here is such that $\bar{\beta} M \SNR = 1$ (i.e., 0 dB), when the expected pathloss $\bar{\beta}$ is calculated for distance $3 d_L$.
Let us assume for simplicity that the coverage area of each RU is a disk with a radius of $3 d_L$, and no random variable in the pathloss model is accounting for shadow fading. For $L=160$, considering that each UE tries to connect only to the RUs with the largest channel gain, we assume a RU-UE pair at distance $d_L$. In that case, the received UL signal of the UE at the RU is approximately $11$ dB above the noise floor for $A=6.25 \times 10^4 \text{ m}^2$, while it is approximately $19$ dB above the noise floor for $A=4 \times 10^6 \text{ m}^2$. The difference of the signal strength between the desired signal and interference signals thus becomes larger for bigger $A$, leading to larger data rates.

Because of the geometry-based UL Tx power, large network areas with fixed $L$ and $K$ can be supported. For all configurations and values of $P_{\rm tx}^{\rm ue}$, the outage probability is below $10^{-4}$. 


\section{Conclusions}
In this paper, we have studied the UL sum SE of cell-free wireless networks with different levels of antenna distribution, user transmit powers and user densities. As explained in many other works, many of the system parameters are interdependent. We have run extensive simulations to show how these dependencies affect the system performance in terms of user outage and sum SE. 
The UE transmit power does not change the sum SE significantly in most systems, but the user outage probability is very much influenced. Thus, an energy efficient approach may be to increase the power level just to the level that no user is in outage. 
The pilot dimension, which limits the number of UEs served by an RU, needs to be increased for a smaller number of RUs, since the average user load per RU becomes larger. Consequently, the optimal user load of a system grows with $\tau_p$. 
With a fixed number of total RU antennas in the system, a certain level of antenna distribution improves the achieved sum SE in general. However, at some point the increased beamforming gain of more concentrated configurations compensates for the smaller distances of associated RU-UE pairs in more distributed configurations. This tradeoff becomes clear in networks with a small area, where more concentrated antenna configurations can outperform the more distributed ones.

\bibliography{IEEEabrv,vtc22-paper}

\end{document}